\documentclass[twocolumn,floatfix,aps,prd,showpacs,amsmath,nofootinbib,
preprintnumbers]{revtex4}

\usepackage{graphicx}

\begin{document}

\preprint{FERMILAB-PUB-05-498-A}

\title{Non-Gaussianity from Broken Symmetries}

\author{Edward W. Kolb}\email{rocky@fnal.gov}
\affiliation{Particle Astrophysics Center, Fermi
        National Accelerator Laboratory, Batavia, Illinois \ 60510-0500, USA \\
        and Department of Astronomy and Astrophysics, Enrico Fermi Institute,
        University of Chicago, Chicago, Illinois \ 60637-1433, USA}

\author{Antonio Riotto}\email{antonio.riotto@pd.infn.it}
\affiliation{CERN Theory Division CH-1211, Geneva 23, Switzerland}

\author{Alberto Vallinotto}\email{avallino@uchicago.edu}
\affiliation{Physics Department, The University of Chicago,
    Chicago, Illinois 60637-1433, USA \\
    and Particle Astrophysics Center, Fermi
    National Accelerator Laboratory, Batavia, Illinois \ 60510-0500, USA}

\date{\today}

\begin{abstract}
Recently we studied inflation models in which the inflaton potential
is characterized by an underlying approximate global symmetry.  In the
first work we pointed out that in such a model curvature perturbations
are generated after the end of the slow-roll phase of inflation. In
this work we develop further the observational implications of the
model and compute the degree of non-Gaussianity predicted in the
scenario.  We find that the corresponding nonlinearity parameter,
$f_{NL}$, can be as large as $10^2$.
\end{abstract}


\maketitle

\section{Introduction \label{sec:INTRO}}

Inflation has become the dominant paradigm to understand the initial
conditions for the density perturbations in the early Universe, which
are the seeds for the Large-Scale Structure (LSS) and for Cosmic
Microwave Background (CMB) temperature anisotropies
\cite{lrreview}. In the inflationary picture, primordial density and
gravity-wave fluctuations are created from quantum fluctuations and
``redshifted'' out of the horizon during an early period of
superluminal expansion of the Universe.

Despite the simplicity of the inflationary concept, the mechanism by
which cosmological curvature (adiabatic) perturbations are generated
is not yet established. In the standard slow-roll inflationary
scenario associated with a single inflaton field, density
perturbations are due to quantum fluctuations of the inflaton
itself. In the curvaton mechanism \cite{curvaton}, the final curvature
perturbation $\zeta$ is produced from an initial isocurvature mode
associated with quantum fluctuations of a light scalar (other than the
inflaton), the curvaton, whose energy density is negligible during
inflation and which decays much after the end of inflation. Recently,
we have proposed an alternative scenario for the generation of the
comoving curvature perturbation
\cite{Kolb:2004jm}.

The idea is that if the inflaton sector is characterized by an {\it
approximate} global symmetry, the associated pseudo-Nambu-Goldstone
boson(s) may be quantum mechanically excited during inflation. This
results in different initial conditions in separate horizon volumes
for the (pre)heating stage after the end of inflation. These different
initial conditions will give rise to fluctuations in the comoving
number densities of the light relativistic states produced during the
decay process and, ultimately, to CMB anisotropies. Other mechanisms
for the generation of cosmological perturbations have been proposed as
well, such as the inhomogeneous reheating scenario \cite{gamma1},
ghost-inflation \cite{ghost}, and the D-celeration scenario
\cite{dacc}, to mention a few.

Given all these theoretical models, it is very important to find ways
to discriminate among them based on observational evidence. A precise
measurement of the spectral index of comoving curvature perturbations
$n_\zeta$ will provide a powerful constraint to slow-roll inflation
models and to the standard scenario for the generation of cosmological
perturbations, which predicts $|n_\zeta-1|$ significantly below
unity. However, alternative mechanisms generally also predict a value
of $n_\zeta$ very close to unity. Thus, even a precise measurement of
the spectral index will not allow us to discriminate efficiently among
them. Also, the lack of gravity-wave signals in CMB anisotropies will
not provide any information about the perturbation generation
mechanism, since alternative mechanisms predict an amplitude of
gravity waves far too small to be detectable by future experiments
aimed at observing the $B$-mode of the CMB polarization.

There is, however, a third observable which will prove fundamental in
providing information about the mechanism chosen by Nature to produce
the structures we see today. It is the deviation from pure Gaussian
statistics, {\it i.e.}, the presence of higher-order connected
correlation functions of CMB anisotropies.  Since for every scenario
there exists a well defined prediction for the strength of
non-Gaussianity and its dependence on the parameters, testing the
non-Gaussian level of primordial fluctuations is one of the most
powerful probes of inflation \cite{review}, and is crucial to
discriminate among different, but otherwise observationally
indistinguishable, mechanisms. For instance, the single-field
slow-roll inflation model produces negligible non-Gaussianity, and its
dominant contribution arises from the evolution of the ubiquitous
second-order perturbations after inflation, which is potentially
detectable with future observations of CMB temperature and
polarization anisotropies. This effect {\it must exist} regardless of
the specific inflationary model considered, setting a lower bound of
the non-Gaussianity of cosmological perturbations. If evidence for
this ubiquitous non-Gaussianity is not found experimentally, the
present understanding of the evolution of cosmological perturbations
will be challenged at a very deep level.

Motivated by the extreme relevance of pursuing non-Gaussianity in the
CMB anisotropies, in this paper we calculate the level of
non-Gaussianity at large scales in the CMB anisotropies produced in
the class of models where the comoving curvature perturbations are
generated by broken symmetries in the inflaton sector
\cite{Kolb:2004jm}. Since we will find that in this class of models
the non-Gaussianity is substantial, as a first approximation it will
be possible to neglect the non-linearity generated by gravitational
dynamics in the post-inflationary phase \cite{bmr,michele} .

The paper is organized as follows. In Sec.\ \ref{sec:Overview} we
present an overview of the general physical mechanism leading to
curvature perturbations production during multi-field inflaton
decay. Sec.\ \ref{sec:Application} then focuses on the determination
of the comoving curvature perturbation produced when the inflaton
field is characterized by a broken U(1) symmetry and when the decay
process is the so-called instant preheating. In Sec.\ \ref{sect:calc},
we carry out the calculation of the non-linearity parameter $f_{NL}$
for the model considered. Sec.\ \ref{sect:GR} then generalizes such
results to determine a general formula for the evaluation of the level
of non-Gaussianity produced whenever the potential is characterized by
a broken symmetry, and summarizes the conclusions.

\section{Model Overview \label{sec:Overview}}

As a first step, let's analyze the consequences of assuming an
inflationary potential characterized by a global broken symmetry for
the production of curvature perturbations during the decay process of
the inflaton field.

In what follows, we consider a multi-field inflationary scenario where
the scale factor $a(t)$ and the background inflaton fields $\phi_i(t)$
evolve according to the system of coupled differential equations
\begin{subequations}
\begin{eqnarray}
    \ddot{\phi}_i+3H\dot{\phi}_i+\frac{\partial V}{\partial
    \phi_i}=0, \quad i=1,...,n,\label{GI:backdyn}\\
    H^2=\left(\frac{\dot{a}}{a}\right)^2=\frac{8\pi}{3 M_p^2}\left[
    \sum_i \frac{\dot{\phi}_i^2}{2}+V(\vec{\phi})\right],\label{GI:H2}
\end{eqnarray}
\end{subequations}
where $\vec{\phi}=(\phi_1,\phi_2,...,\phi_n)$.

As in the standard single-field scenario, the multi-field scenario is
characterized by an initial slow-roll phase during which quantum
fluctuations of all the light degrees of freedom are excited,
stretched to macroscopic scales, and frozen in amplitude once their
wavelength exceeds the Hubble radius. It is important to note that in
the single-field scenario only fluctuations parallel to the direction
of motion in field space could be generated. As was shown in Refs.\
\cite{Kolb:2004jm,Gordon:2000hv}, these fluctuations would
later correspond to \textit{adiabatic} perturbations. In the
multi-field case, on the other hand, fluctuations both parallel
\textit{and} orthogonal to the direction of motion in field space
can be excited, the latter corresponding to \textit{isocurvature}
perturbations. In general, then, at the end of the slow roll phase
the value of the background inflaton field will have acquired a
spatial dependence
\begin{equation}\label{GI:deltaphi}
    \vec{\phi}(t_0,x)=\vec{\phi}_0(t_0)+\delta\vec{\phi}(t_0,x),
\end{equation}
where $t_0$ denotes the epoch after termination of the slow-roll phase
but before the decay process. It is necessary
to point out that up to this point in the evolution history of the
Universe the only difference between the single-field and the
multi-field scenarios consists of the fact that field perturbations
in the former case are of the adiabatic type only, while in the
latter case they consist of both adiabatic \textit{and} isocurvature
perturbations.

Moving on to the next phase in the evolution history of the Universe,
it is very important to note that the quantum fluctuations imprinted
onto the value of the inflaton fields during the slow-roll phase also
represent, through Eq.\ (\ref{GI:deltaphi}), perturbations in the
\textit{initial conditions} for the evolution of the background fields
$\vec{\phi}$ during the decay phase.  Similar to a chaotic mechanical
system, \textit{if} the potential $V(\vec{\phi})$ is characterized by
a global broken symmetry, the system of differential equations Eqs.\
(\ref{GI:backdyn},\ref{GI:H2}) then exhibits a sensitive dependence on
the initial conditions. This fact, in turn, guarantees that different
initial conditions will correspond to evolution histories for the
background fields $\vec{\phi}(t)$ for $t\ge t_0$ which do not just
differ from one another by a simple time translation. Then to
summarize, the broken symmetry of the potential is sufficient to
guarantee that the field perturbations generated during the slow-roll
phase will correspond to field trajectories $\vec{\phi}(t)$ that will
differ substantially in the subsequent phase of inflaton decay.

Moving on to consider the decay process, it is clear that if the
inflaton fields $\vec{\phi}$ have to decay into some other particles
$\chi$, then the two must necessarily be connected by an interaction
Lagrangian $\mathcal{L}_{\chi\phi}$. Furthermore, this interaction
Lagrangian necessarily must depend on the value of the fields
involved, $\vec{\phi}$ and $\chi$. Denoting by $n_{\chi}$ the comoving
number density of $\chi$ particles produced during the decay process,
it is then trivial to note that such a quantity will be a (usually
very complicated) functional of the evolution histories of the
background fields. But since the field trajectories will depend on the
initial conditions set at time $t_0$, then formally we may write
\begin{equation}
    n_{\chi}=F[\vec{\phi}(t_0),...].
\end{equation}

In a multi-field scenario the perturbations generated during the
slow-roll phase will appear as perturbations of the initial conditions
for the evolution of the background fields during the decay
stage. This will in turn result in fluctuations in the comoving number
density of particles produced
\begin{equation}
    \delta\vec{\phi}(t_0, x)\Rightarrow \delta n_{\chi}(t_1,x),
\end{equation}
where $t_1>t_0$ denotes the end of the decay process.

To determine the curvature perturbations produced during such a decay
process, it is sufficient to note that the generated energy density
is proportional to the comoving number density of produced particles,
$\rho_{\chi}\sim n_{\chi}$. Assuming for simplicity complete decay,
an estimate of the curvature perturbation $\zeta$
generated during this phase is then
\begin{equation}\label{GI:zeta}
    \zeta \equiv -\psi-H\frac{\delta\rho_{\chi}}{\dot{\rho}_{\chi}}
     \approx \alpha\frac{\delta n_{\chi}}{n_{\chi}},
\end{equation}
where in the last step the \textit{spatially flat gauge} has been
assumed and $\alpha$ is a proportionality constant whose numerical
value depends on the redshifting of the particle produced.

The level of the non-Gaussianity produced is usually specified using
the non-linearity parameter $f_{NL}$, which determines the
non-Gaussian contribution to the Bardeen potential $\Phi$ according to
\begin{equation}\label{GI:fnldef}
    \Phi=\Phi_G+f_{NL}\Phi_G^2,
\end{equation}
where $\Phi_G$ is the Gaussian part of the Bardeen potential. It is
then possible to connect $\Phi$ to $\zeta$ through\footnote{It is
important to note that Eq.\ (\ref{GI:Phi}) connecting $\Phi$ and
$\zeta$ is valid only at linear order. It is nonetheless correct to
use it provided that the value of the non-Gaussianity parameter is
larger than unity, which will be the case in what follows.}
\begin{equation}
\label{GI:Phi}
    \Phi=-\frac{3}{5}\,\zeta=-\frac{3\alpha}{5}
    \frac{\delta n_{\chi}}{n_{\chi}},
\end{equation}
where in the last step the expression for the curvature perturbation
$\zeta$ produced in the present model has been used.

It seems important to stress here that the \textit{only} crucial
assumption made so far concerns the global broken symmetry of the
potential. If the symmetry is unbroken, it is possible to see that the
perturbations of the initial conditions would produce evolution
histories of the background fields $\vec{\phi}(t)$ which would only
differ by a time translation: the curvature perturbations thus
produced would simply represent a gauge artifact and could be gauged
away by a suitable choice of the slicing and threading. The only other
assumption made is that the inflaton decays into the $\chi$ field
through the non-perturbative process of preheating, but this
assumption is not crucial to the final conclusions that fluctuations
in the initial conditions at the beginning of the decay stage would
yield fluctuations in the energy density and therefore in the
curvature.

Curvature perturbations produced through this mechanism can be
characterized by significant degree of non-Gaussianity. The next
section contains a simple example which shows how it is possible to
determine the value of the non-Gaussianity parameter $f_{NL}$ in a
specific case.

\section{An Analytic Example: The Broken U(1) Case in the Instant
Preheating Scenario}\label{sec:Application}

Because of the general characteristics of the process being considered
-- a sensitive dependence of the field-space trajectories on the
initial conditions and the non-perturbative nature of the preheating
process -- the general evaluation of the curvature perturbation and of
the non-linearity parameter generated usually require a numerical
approach. In this section the \textit{instant preheating} model of
Felder {\it et al.}\ \cite{Felder:1998vq} is considered in order to
obtain an analytic estimate of $f_{NL}$. It is important to stress,
however, that while the choice of the specific preheating model is in
part dictated by computational convenience, the results of the present
section can be easily generalized to take into account different decay
processes and background-field spaces.

To proceed further it is necessary to specify two details that so far
have been left undetermined: the characteristics of the
background-field space (dimensionality, potential, and broken
symmetry) and the details of the preheating mechanism. The former
determines the field trajectories $\vec{\phi}(t)$, while the latter
allows the specification of the functional relating the trajectories
to the comoving density of particles produced,
$n_{\chi}[\vec{\phi}(t)]$.

\subsubsection{The Choice of the Background Field Landscape}

Proceeding as in Ref.\ \cite{Kolb:2004jm}, the dimensionality of the
field space is chosen to be 2, so that the inflaton-field landscape is
described by $\phi_1$ and $\phi_2$. This choice reflects the fact that
curvature perturbations and possibly non-Gaussianity can arise in the
simplest multi-field inflationary scenario. The complex field $\phi$
can then be introduced:
\begin{equation}
    \phi=\phi_1+i\phi_2=|\phi|e^{i\theta}.
\end{equation}

The potential $V(\phi_1,\phi_2)$, which is characterized by a slightly
broken U(1) symmetry, can be modeled close to its minimum by
\begin{equation}\label{App:V}
V(\phi_1,\phi_2)=\frac{m^2}{2}\left[\phi_1^2+\frac{\phi_2^2}{(1+x)}\right],
\end{equation}
where $x$ represents a measure of the symmetry breaking. This choice
for the potential landscape allows one to determine analytically the
field trajectories $\vec{\phi}(t)$ if the Hubble expansion is
neglected for the short time period during which the decay process
occurs (see the Appendix of Ref.\ \cite{Kolb:2004jm}).

Finally, the initial conditions for the background dynamics set at the
end of the slow-roll phase -- and fluctuations thereof -- can be
equivalently expressed as $[\phi_1(t_0),\phi_2(t_0)]$ or as
$[|\phi|(t_0), \theta(t_0)]$. However, since the aim of the present
work is to analyze the conversion of isocurvature perturbation into
adiabatic perturbations during the decay stage, and since the form of
the potential is such that the motion of the field is almost
completely in the radial direction, in what follows the latter choice
will be adopted together with the shorthand notation $[|\phi_0|,
\theta_0]$.

\subsubsection{The Choice of the Inflaton Decay Process}

In this analysis, the \textit{instant preheating} model of Felder
\textit{et al.}\ \cite{Felder:1998vq} is chosen as the
process through which the inflaton decays into lighter particles.
Such a choice is made both because it is not unreasonable to think
that the preheat field $\chi$ is coupled to other fields into which
it can perturbatively decay, and because it also yields
a convenient analytical expression for the functional
$n_{\chi}[\vec{\phi}(t)]$.

The complex inflaton field $\phi$ is coupled to the preheat field
$\chi$ through the standard preheating interaction Lagrangian
$\mathcal{L}_{\chi\phi}=-\frac{1}{2}g|\phi|^2\chi^2$. The preheat
field $\chi$ is also coupled to a massless fermion field $\psi$ by the
interaction Lagrangian $\mathcal{L}_{\phi\psi}=h\bar{\psi}\psi\chi$,
so that the perturbative decay $\chi\rightarrow\bar{\psi}\psi$ can
take place.\footnote{Given the large mass of the other fields involved
in the decay process, the assumption of a massless $\psi$ is actually
immaterial.} Applying the results of Kofman \textit{et al}.\
\cite{Kofman:1997yn} and of Felder \textit{et al}.\
\cite{Felder:1998vq}, the comoving number density of $\chi$
particles produced during the first half-oscillation of $\phi$ close
to its minimum can be determined to be
\begin{equation}\label{App:nchi}
    n_{\chi}=\frac{\left(g
    |\dot{\phi}_*|\right)^{3/2}}{8\pi^3}\exp\left[ -\frac{\pi
    g|\phi_*|^2}{|\dot{\phi}_*|}\right] ,
\end{equation}
where $t_*$ is the instant when the inflaton $\phi$ reaches the
minimum of the potential $V(\phi)$ \textit{along its trajectory} and
$|\dot{\phi}_*|$ and $|\phi_*|$ represent the field velocity and
distance from the absolute minimum of the potential at such an
instant. From the interaction Lagrangian $\mathcal{L}_{\chi\phi}$ it
is possible to note that as the inflaton moves away from the bottom of
the potential, the preheat field is endowed with an increasing
effective mass $m_{\chi}=g|\phi|$. The $\chi$ particles thus produced
then acquire a large mass and decay into $\psi$ particles with a decay
rate $\Gamma_{\chi\rightarrow\bar{\psi}\psi}=h^2m_{\chi}/8\pi=h^2g|\phi|/8\pi$
that is proportional to $|\phi|$, which is increasing as the
background inflaton moves away from the bottom of the potential.
Depending on the values of the coupling constants, the whole decay
chain $\phi\rightarrow\chi\rightarrow\psi$ can proceed very
efficiently, turning \textit{all} the energy initially stored in the
$\phi$ fields into the fermions $\psi$ in a single half-oscillation of
the inflaton around the minimum of its potential. The fermions then
rapidly thermalize and complete the reheating process.

The fluctuations in the final energy density (and thus in the CMB
temperature) can then be traced back to the fluctuations in the number
density of preheat particles produced during the decay process, which
are in turn caused by the isocurvature perturbations generated during
the slow-roll phase.

\section{The Calculation of the Non-Linearity Parameter}\label{sect:calc}

To obtain an estimate of the curvature perturbation through Eq.\
(\ref{GI:zeta}), it is necessary to find an expression for $\delta
n_{\chi}/n_{\chi}$. From Eq.\ (\ref{App:nchi}) it is straightforward
to get \cite{Kolb:2004jm}:
\begin{equation}
    \frac{\delta n_{\chi}}{n_{\chi}}=\left(\frac{3}{2}+
    \frac{\pi g |\phi_*|^2}{|\dot{\phi}_*|}
    \right)\frac{\delta |\dot{\phi}_*|}{|\dot{\phi}_*|}-
    \frac{2\pi g |\phi_*|^2}{|\dot{\phi}_*|}\frac{\delta
    |\phi_*|}{|\phi_*|}.
\end{equation}

Moving on to consider the background-field trajectories, it is
possible to note that the choice of the potential, Eq.\ (\ref{App:V}),
allows one to derive the following approximate values for $|\phi_*|$
and $|\dot{\phi}_*|$ as functions of the initial conditions
$[|\phi_0|,\theta_0]$ and of the value of the symmetry breaking
parameter $x$ (see the Appendix of Ref.\ \cite{Kolb:2004jm}):
\begin{subequations}
\begin{eqnarray}
    |\phi_*(|\phi_0|,\theta_0;x)|&\approx& \frac{|\phi_0| \pi x}{2\sqrt{2}}
    |\sin(2\theta_0)| ,
    \label{App:lapprox}\\
    |\dot{\phi}_*(|\phi_0|,\theta_0;x)|&\approx&
    m|\phi_0|\sqrt{1-x\sin^2(\theta_0)
    \label{App:vapprox}}.
\end{eqnarray}
\end{subequations}
Let's then note that the form of the potential, Eq.\ (\ref{App:V}), is
such that the motion of the inflaton is almost completely in the
radial direction. Perturbations in the radial direction therefore
correspond to adiabatic perturbations, while perturbations in the
azimuthal direction -- leaving the value of the potential unaltered --
correspond to isocurvature perturbations.\footnote{For a more thorough
analysis of this point, involving the fact that the field trajectory
defines a local coordinate system in field space, we refer the reader
to the discussion given in Ref.\ \cite{Kolb:2004jm}.}  Since the aim
of the present work is to analyze the conversion of isocurvature
perturbations into adiabatic perturbations and the resulting curvature
perturbations, let's then assume that the slow-roll phase of inflation
was such that\footnote{This happens if the potential in the slow-roll
region is such that the effective mass for excitation of quantum
fluctuations of the inflaton in the direction of motion is larger than
the Hubble parameter during slow-roll.}
\begin{equation}
    \delta\theta_0\gg\delta|\phi_0|.
\end{equation}
To obtain an estimate of the value of the non-Gaussianity parameter
$f_{NL}$, the Bardeen potential $\Phi$ needs to be expressed as
\begin{equation}
    \Phi=f_1(\theta_0)\delta\theta_0+f_2(\theta_0)(\delta\theta_0)^2.
\end{equation}
Then, recalling Eq.\ (\ref{GI:fnldef}), it is possible to identify
\begin{subequations}
\begin{eqnarray}
    \Phi_G&=&f_1(\theta_0)\,\delta\theta_0,\\
    f_{NL}&=&\frac{f_2(\theta_0)(\delta\theta_0)^2}{\Phi_G^2}
    =\frac{f_2(\theta_0)}{f_1^2(\theta_0)}.
\end{eqnarray}
\end{subequations}
It is also necessary to point out that while in Ref.\
\cite{Kolb:2004jm} we performed the calculation also including terms
of order $x^2$ and $x^3$, in what follows only terms linear in the
symmetry breaking parameter will be considered. Let's then notice that
the term
\begin{equation}
    \frac{\pi g |\phi_*|^2}{|\dot{\phi}_*|}=\frac{\pi^3 g x^2 |\phi_0|
    \sin^2(2\theta_0)}{8m\sqrt{1-x\sin^2(\theta_0)}},
\end{equation}
is of order $x^2$, which then means that every term that contains it
as a multiplying factor can safely be neglected at linear order in
$x$. This then necessarily implies that
\begin{equation}
    \Phi\approx-\frac{9\alpha}{10}\frac{\delta
    |\dot{\phi}_*|}{|\dot{\phi}_*|}.
\end{equation}
From Eqs.\ (\ref{App:lapprox},\ref{App:vapprox}) it is then possible
to derive the second order expressions for
\begin{subequations}
\begin{eqnarray}
    &&\frac{\delta |\phi_*|}{|\phi_*|}= 2\frac{\cos(2\theta_0)}
         {\sin(2\theta_0)}
    \delta\theta_0-\delta\theta_0^2,\\
    &&\frac{\delta
    |\dot{\phi}_*|}{|\dot{\phi}_*|}=
    -\frac{x}{2\left[1-x\sin^2(\theta_0)\right]}
    \left\{\sin(2\theta_0)\delta\theta
    \phantom{\frac{\sin^2}{\left[\sin^2\right]}}
     \right. \nonumber \\
    & & \left. +\left[\cos(2\theta_0)+\frac{x\sin^2(2\theta_0)}
    {4\left[1-x\sin^2(\theta_0)\right]}\right]\delta\theta_0^2\right\} ,
\end{eqnarray}
\end{subequations}
where the expression for $\delta |\phi_*|/|\phi_*|$ has been added
for completeness. We then have
\begin{eqnarray}
    &&\Phi=\frac{9\alpha x}{20\left[1-x\sin^2(\theta_0)\right]}
    \left\{\sin(2\theta_0)\delta\theta
        \phantom{\frac{\sin^2}{\left[\sin^2\right]}}
    \right. \nonumber \\ && \left.+\left[\cos(2\theta_0)
    +\frac{x\sin^2(2\theta_0)}
    {4\left[1-x\sin^2(\theta_0)\right]}\right]\delta\theta_0^2\right\},
\end{eqnarray}
which then leads to the identification
\begin{subequations}
\begin{eqnarray}
    f_1(\theta_0)&=&\frac{9\alpha x}{20\left[1-x\sin^2(\theta_0)\right]}
    \sin(2\theta_0),\\
    f_2(\theta_0)&=&\frac{9\alpha x}{20\left[1-x\sin^2(\theta_0)\right]}
    \nonumber\\
    &\times&\left[\cos(2\theta_0)+\frac{x\sin^2(2\theta_0)}
    {4\left[1-x\sin^2(\theta_0)\right]}\right].
\end{eqnarray}
\end{subequations}
It is then immediate to find that
\begin{eqnarray}\label{App:fnl}
    f_{NL}&=&\frac{20}{9\alpha}\frac{\left[1-x\sin^2(\theta_0)\right]}
    {x\sin^2(2\theta_0)}\nonumber \\
    & & \times\left[\cos(2\theta_0)+\frac{x}{4}
    \frac{\sin^2(2\theta_0)}{\left[1-x\sin^2(\theta_0)\right]}\right]
    \nonumber\\
    &\approx&\frac{20}{9\alpha}\frac{\cos(2\theta_0)}{x\sin^2(2\theta_0)},
\end{eqnarray}
where in the last expression only the leading term has been
considered.\footnote{It is in fact important to note that if we wanted
to perform a calculation exact up to $x^0$ order we should also
consider a term arising from the $\delta|\phi_*|/|\phi_*|$ factor.} It
is then very interesting to note this particular model is
characterized by a value of the non-linearity parameter which is
inversely proportional to the value of the symmetry breaking: in
general slightly broken symmetries should then produce mild curvature
perturbations characterized by large non-Gaussianity.

\begin{figure}
\includegraphics[width=0.48\textwidth]{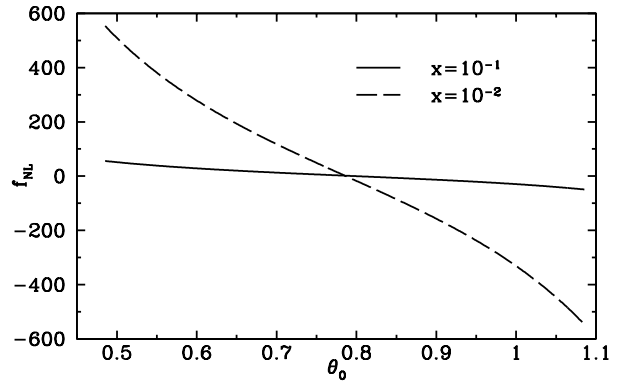}
\caption{\label{Fig:1}Value of $f_{NL}$ for values of the symmetry
breaking parameter of $x=10^{-1}$ (solid curve) and $x=10^{-2}$
(dashed curve). It seems important to stress that only a very small
interval in the initial conditions $\theta_0$ will yield
\textit{insignificant} non-Gaussianity.}
\end{figure}

\section{General Result for $f_{NL}$ and Conclusions\label{sect:GR}}

The results of the previous section were obtained under very specific
assumptions: a broken $U(1)$ symmetry and the \textit{instant
preheating} mechanism. It is nonetheless possible to generalize the
previous approach to obtain a general formula to estimate the value of
the non-linearity parameter of the curvature perturbations produced in
a multi-field scenario by a generic decay process.

As it was shown in Ref.\ \cite{Kolb:2004jm}, the initial conditions
perturbation $\delta\vec{\phi}(t_0,x)$ can in fact be decomposed into
an adiabatic component $\delta\phi_{\parallel}$ parallel to the field
velocity $\hat{u}_{\parallel}\equiv d\vec{\phi}/dt$ and into an
isocurvature component $\delta\phi_{\perp}$ perpendicular to it and
directed along the unit vector $\hat{u}_{\perp}$. As in Ref.\
\cite{Kolb:2004jm}, it is assumed here as well that the adiabatic
perturbations generated during the slow-roll phase of inflation are
actually negligible compared to the isocurvature component
\begin{equation}
    \delta\phi_{\perp}\gg\delta\phi_{\parallel}.
\end{equation}
This immediately leads to the following estimate for the Gaussian
curvature perturbation
\begin{equation}
    \Phi_G\approx -\frac{3\alpha}{5 n_{\chi}}
    \frac{\partial n_{\chi}}{\partial
    \phi_{\perp}}\delta\phi_{\perp}.
\end{equation}
It is important to stress that the functional relating the comoving
number density of particles produced during the decay process to the
background trajectory is in general very complicated, usually
requiring numerical solutions to formulate quantitative predictions.
Furthermore, if the potential is characterized by a broken symmetry
then the background field trajectories may also be characterized by
a sensitive dependence on the initial conditions and therefore may
exhibit chaotic behavior. All these factors then conspire to make
the evaluation of $\Phi$ rather involved.

Recalling Eqs.\ (\ref{GI:fnldef}, \ref{GI:Phi}) and expanding $\delta
n_{\chi}/n_{\chi}$ to second order and neglecting the terms containing
powers of $\delta\phi_{\parallel}$, the Bardeen potential turns out to
be
\begin{equation}
    \Phi=-\frac{3 \alpha}{5 n_{\chi}}
    \left[
    \frac{\partial n_{\chi}}{\partial \phi_{\perp}}\delta\phi_{\perp}+
    \frac{1}{2}\frac{\partial^2 n_{\chi}}
    {\partial \phi_{\perp}^2}(\delta\phi_{\perp})^2
    \right],
\end{equation}
which then allows one to obtain the following general expression for the
non-linearity parameter
\begin{equation}\label{GI:fnl}
    f_{NL}=-\frac{5 n_{\chi}}{6\alpha}
    \frac{\partial^2 n_{\chi}/\partial\phi_{\perp}^2}
    {\left(\partial n_{\chi}/\partial \phi_{\perp}\right)^2}.
\end{equation}

It is then clear that depending on the relative magnitudes of the
derivatives of $n_{\chi}$, significant or negligible non-linearity
can be produced by a multi-field potential characterized by a global
broken symmetry. In particular, it is straightforward to obtain the
following condition
\begin{equation}\label{GI:conditionfnl1}
    f_{NL}\ge 1 \Leftrightarrow \frac{1}{n_{\chi}}\frac{\partial^2 n_{\chi}}
    {\partial \phi_{\perp}^2}\ge\frac{2\alpha}{n_{\chi}^2}
    \left(\frac{\partial n_{\chi}}{\partial \phi_{\perp}}\right)^2.
\end{equation}
which, given the chaotic nature of the background field trajectories
and the exponential production of particles in the resonances
characteristics of the preheating processes, can possibly be
satisfied by a large class of models.

A few comments are finally in order. The level of non-Gaussianity
predicted by the model at hand can be rather sizable, values of
$f_{NL}\sim 100$ can be easily achieved. As we have already mentioned
in the Introduction, in the computation of the nonlinearity parameter
$f_{NL}$ we have neglected the nonlinear gravitational contribution
coming from the post-inflationary evolution. Since they amount to a
contribution to $f_{NL}$ of order unity \cite{review}, it follows that
the estimate of $f_{NL}$ will be valid only in the regime $f_{NL}\gg
1$. In other words, the validity of the estimate requires that the
global symmetry is slightly broken, that is $x\ll 1$. The fact that
the level of non-Gaussianity is inversely proportional to the
parameter $x$, which gives a measure of the breaking of the global
symmetry within the inflaton sector, is reminiscent of what happens
for the curvaton scenario \cite{ung}. There the non-Gaussianity is
inversely proportional to the parameter $r$ which measures the energy
density contribution of the curvaton to the total one at the time of
curvaton decay. Since present WMAP data impose a bound
$\left|f_{NL}\right|\lesssim 100$, this translates into a weak bound
on the breaking parameter
\begin{equation}
x\gtrsim 10^{-2} \, \tan \left(2\theta_0\right)\sin
\left(2\theta_0\right).
\end{equation}
Since future tests of the angular bispectrum of temperature anisotropy
alone could detect non-Gaussianity as small as
$\left|f_{NL}\right|\sim 5$ \cite{review}, it would be interesting to
generalize the computation performed in this paper to include the
contribution of gravitational nonlinearities which render the
nonlinearity parameter $f_{NL}$ momentum-dependent. As explicitly
shown in Ref. \cite{michele}, this dependence augments the
non-Gaussian signal at large multipoles and may be therefore crucial
in testing large symmetry breaking patterns in the inflaton sector.

\acknowledgments
A.R,\ is on leave of absence from INFN, Padova (Italy). E.W.K.\ and A.V.\ were
supported in part by NASA grant NAG5-10842 and by the Department of Energy.



\end{document}